\documentclass[12pt]{article}

\usepackage{graphics,psfrag}
\usepackage{amsthm,amssymb,epsfig,amsmath,euscript,array,cite}
\usepackage{color}
\usepackage{graphicx}
\usepackage[bulletsep]{collref}
\newcommand{\be}{\begin{equation}}
\newcommand{\ee}{\end{equation}}
\def\bea{\begin{eqnarray}}
\def\eea{\end{eqnarray}}
\newcommand{\eq}[1]{(\ref{#1})}
\def\nn{\nonumber}

\def\gh{{\hat\gamma}}

\newcommand{\beq}{\begin{equation}}
\newcommand{\eeq}{\end{equation}}
\newcommand{\ben}{\begin{eqnarray}}
\newcommand{\een}{\end{eqnarray}}
\newcommand{\bes}{\begin{subequations}}
\newcommand{\ees}{\end{subequations}}
\newcommand{\blg}{\begin{align}}
\newcommand{\elg}{\end{align}}

\newcommand{\Tr}{{\rm Tr}}
\newcommand{\tr}{{\rm tr}}
\newcommand{\cN}{{\cal N}}

\newcommand{\prt}[1]{{\left( {#1} \right)}}

\newcommand{\startappendix}{
\setcounter{section}{0}
\renewcommand{\thesection}{\Alph{section}}}

\def\Tr{{\rm Tr}}
\def\one{\mbox{1 \kern-.59em {\rm l}}}

\def\a{\alpha}      
\def\b{\beta}       
  \def\G{\Gamma}  
  \def\D{\Delta}  
\def\e{\epsilon}

\def\l{\lambda} 
\def\m{\mu} \def\n{\nu}
\def\o{\omega}
 \def\P{\Pi}

\def\s{\sigma}  
\def\t{\tau}
\def\th{\theta}

 \def\cN{{\cal N}}

\begin{document}

\hfill{}

\vspace{14pt}

\begin{center}

{\Large \bf Generalised cusp anomalous dimension in \\
$\beta-$ deformed super Yang Mills theory

}
\vspace{18pt}

{\bf
 George Georgiou$^1$, Dimitrios Giataganas$^{2,3}$
}

{\em $^1$ Demokritos National Research Center \\ Institute of Nuclear and Particle Physics\\
Ag. Paraskevi, GR-15310 Athens, Greece\\~\newline
$^2$Department of Engineering Sciences, \\University of Patras,
26110 Patras, Greece\\~\newline
$^3$National Institute for Theoretical Physics,\\
School of Physics and Centre for Theoretical Physics,\\
University of the Witwatersrand,\\
Wits, 2050,
South Africa
}

{\small \sffamily
georgiou@inp.demokritos.gr~,~dimitrios.giataganas@gmail.com
}

\vspace{18pt}
{\bf Abstract}\end{center}
In this work we study the cusp anomalous dimension of the marginally deformed ${\cal N}=4$ sYM theory. We find the expression of the cusp anomalous dimension both at the weak and strong coupling limits.  On the gravity side we partially map the system of equations to that of undeformed $\cN=4$ sYM, by defining in the internal space an effective angle $\Delta \theta_{eff}$ which depends on the deformation parameter. The cusp anomalous dimension then can be read from that of the $\cN=4$ sYM theory, by using this effective angle instead of the usual angle of the undeformed 5-sphere. In the weak coupling regime we find no $\beta$ dependence up to two loops. Based on these results we conjecture that for any value of $\lambda$ the cusp anomalous dimension is given by the $\cN=4$ sYM result, using the coupling dependent effective angle. As a consequence of this, we derive a BPS-like condition between the angle of the $AdS$ space and the $\Delta \theta_{eff}$ of the deformed sphere which when satisfied, nullifies the cusp anomalous dimension.

\setcounter{page}0
\newpage

\section{Introduction}
The AdS/CFT correspondence~\cite{Maldacena:1998re,Witten:1998qj} claims a duality between the maximally supersymmetric
field theory in four dimensions, the
${\cal N}=4$ Super Yang-Mills  (sYM) theory and the type-IIB superstring theory on $AdS_5 \times S^5$ background. Remarkably, both theories appear to be integrable (for a recent review see \cite{Beisert:2010jr}), at least at the planar level, while signs of integrability appear in some special large $N$ but non-planar limits, e.g. \cite{Koch:2011hb,Carlson:2011hy}.
The presence of integrability allows oneself to hope that one day both theories will be "solved" and the AdS/CFT correspondence will be strictly proven. On the field theory side it is needed to specify the conformal dimensions of all composite operators of the theory, as well as
the structure constants that determine the Operator Product Expansion (OPE) between two primary operators. In the planer limit, using properties of the integrability several complications were overcome, for example in the study of spectrum of local operators.
In some cases it is even possible to find a non-trivial interpolating function between weak and strong coupling,  for any value of the coupling constant. On the other hand, less is known about the structure constants. For some progress in this direction see
\cite{Escobedo:2010xs,Escobedo:2011xw,Gromov:2012uv} and \cite{Georgiou:2012zj,Georgiou:2011qk,Georgiou:2010an,Georgiou:2008vk} and references therein.

Besides the prototype of the AdS/CFT correspondence there are other cases where a four-dimensional gauge theory is conjectured to
be equivalent to a certain string theory on some supergravity background.
One of those cases is the duality obtained from the original Maldacena's conjecture by marginal deformations. The field theory is obtained from the Lagrangian of ${\cal N} = 4$ sYM, deformed by exactly
marginal operators \cite{Leigh:1995ep} which preserves the $\cN=1$ supersymmetry and has at least a global $U\prt{1}\times U\prt{1}$ symmetry. The gravity dual theory may be obtained by applying a solution generating $SL(3,R)$ transformation to the initial background, or equivalently by performing a TsT transformation on the original metric. This consists of a T-duality along a $U\prt{1}$ direction, then a shift on an another $U\prt{1}$ direction where the deformation parameter enters and finally a T-duality along the initial $U\prt{1}$ symmetry \cite{Lunin:2005jy,Frolov:2005dj}.

This duality was thoroughly studied in the literature.  Papers which focused on perturbative effects in  $\beta$-deformed gauge
theories include \cite{Freedman:2005cg,Mauri:2005pa,Mauri:2006uw,Fiamberti:2008sm}. The amplitudes in the $\beta$-deformed conformal Yang-Mills were considered in \cite{Khoze:2005nd}. Giant magnons configurations were examined in
\cite{Chu:2006ae,Bykov:2008bj} while instanton effects were studied in \cite{Georgiou:2006np,Durnford:2006nb}.
Wilson loop configurations were examined in \cite{Chu:2007pb} and in higher representations in \cite{Imeroni:2006rb}.
The correspondence was also analyzed from the perspective of integrability in \cite{Berenstein:2004ys, Beisert:2005if,Frolov:2005ty,Frolov:2005dj,Gromov:2010dy,Arutyunov:2010gu,deLeeuw:2010ed,Zoubos:2010kh} while, more recently and under the assumption of integrability,
the worldsheet S-matrix of $\beta$-deformed sYM has been proposed in
\cite{Ahn:2010ws,Ahn:2012hs}.

One of the most interesting natural observables in the gauge theories is the Wilson loop operators realized in the dual theory as infinite open strings.
In the maximally supersymmetric field theory in four dimensions ${\cal N}=4$ sYM,
the proper definition of the Wilson loop operator involves the scalar fields \cite{Maldacena:1998im}. More precisely,
\begin{eqnarray}\label{Wilsonloop0}
&&W[C]=\frac{1}{N}\tr P \exp \prt{i g\oint (A_{\mu} \dot{x^{\mu}}\prt{\t}-i \dot{y}^i\prt{\t} \phi_i)d\t},\\\nn
&&\quad\mbox{with}\quad \dot{y_i}=n^i |\dot{x}|,\quad\sum_{i=1}^6 n_i^2=1~.
\end{eqnarray}
An other interesting quantity related to the Wilson loops is  the cusp anomalous dimension. It is defined as the coefficient of the
logarithmic divergence developed by a Wilson loop
operator when there is a cusp of angle $\varphi$ in its contour,
i.e. $\langle W\rangle\sim \exp{(-\Gamma_{cusp}(\varphi,g)\log\Lambda_{IR}/\Lambda_{UV})} $, where $\Lambda_{IR}$ and $\Lambda_{UV}$ are IR and UV cutoffs, respectively.
Furthermore, one can define a
generalized cusp anomalous dimensions $\Gamma_{cusp}(\varphi,\theta,g)$ since  the direction ${\vec n}$ need not be constant but
can abruptly change at the cusp. That is we can define two different points on the five-sphere ${\vec n}_1$ and ${\vec n}_2$, where ${\vec n}_1$ and ${\vec n}_2$ are the directions before and after the cusp. As a result, one can have an angle $\theta$ in the internal space given by $\cos \theta={\vec n}_1\cdot {\vec n}_2$.
In the case where the angle in space-time is equal to the angle in the internal space $\varphi=\pm \theta$ the configuration is
1/4 BPS and therefore the expectation value of the Wilson loop is protected resulting to a vanishing cusp anomalous dimension \cite{Zarembo:2002an}.
The generalized cusp anomalous dimension of ${\cal N}=4$ sYM was calculated at one-loop order, as well as at strong coupling by means of AdS/CFT in
\cite{Drukker:1999zq}.

As stressed in \cite{Correa:2012at} the cusp anomalous dimension is related to variety of other physical observables.
Namely, the coefficient of the large $\varphi$ expansion of the cusp anomalous dimension not only governs the IR divergences in the scattering of massless particles but it is also related to the anomalous dimensions of twist-2 operators. Furthermore, $\Gamma_{cusp}$ characterizes the IR divergences in the scattering of massive particles in the planar limit.
Most importantly, it can be identified with the static quark/anti-quark potential when the two particles sit at an angle $\pi-\varphi$
on a three-sphere $S^3$. This can be easily seen since by the plane to cylinder map, one can map the cusped contour of Figure 1
to two anti-parallel lines on the sphere. Finally, the coefficient of the small $\varphi$ expansion of the cusp anomalous dimension
was calculated to all-loops and for any number of colors in ${\cal N}=4$ sYM \cite{Correa:2012at} and was related to the energy
radiated by a moving quark. This all loop expression was found to be in agreement with the result obtained from a TBA system of
integral equations for $\Gamma_{cusp}(\varphi,\theta,g)$ obtained in \cite{Correa:2012hh,Drukker:2012de}. The two and three-loop expressions for the cusp anomalous dimension were obtained in  \cite{Makeenko:2006ds} and
\cite{Correa:2012nk}, respectively. Finally, the quark/anti-quark potential at weak and strong coupling was also studied in
\cite{Drukker:2011za,Forini:2010ek,Correa:2012at}.

Motivated by these results we focus on this important quantity in the context of the correspondence between the marginally deformed ${\cal N}=1$ sYM theory and its gravity dual. In Section 2 we give a very brief review of the $\beta$-deformed sYM and its gravity dual. In Section 3 we calculate the generalized cusp anomalous dimension in the strongly coupled
$\beta$-deformed gauge theory. We restrict ourselves to the case where the string world-sheet lives on a deformed two-sphere ${\tilde S}^2\subset {\tilde S}^5$. This means that the corresponding Wilson loop operator couples to only three of the six scalar fields of field theory.
By finding the minimal surface solution we see that the generalized cusp anomalous dimension, or equivalently the quark/antiquark potential,
explicitly depends on the deformation parameter $\gh$ and can be written in a form closely related to that of $\cN=4$ sYM.
More precisely,
\begin{eqnarray}\label{cuspstrong0}
 \Gamma_{cusp}^{\beta}(\varphi,\Delta\theta,\Delta\phi_1,\lambda,\beta)= \Gamma_{cusp}^{N=4}(\varphi,\Delta\theta_{eff},\lambda)+{\cal O}(\lambda^0),
\end{eqnarray}
where
\begin{eqnarray}\label{zxzzz}
\Delta\theta_{eff}=\Delta\theta_{eff} (\Delta\theta,\Delta\phi_1,\beta\sqrt{\lambda})~,
\end{eqnarray}
is a function of the angles in the internal space and of the deformation parameter.
Equation \eqref{cuspstrong0} implies that the cusp anomalous dimension vanishes when
\begin{eqnarray}\label{BPS-cond0}
\varphi=\Delta\theta_{eff}(\Delta\theta,\Delta\phi_1,\beta\sqrt{\lambda})~,
\end{eqnarray}
for large $N$. In Section 3.1 we determine $\Delta\theta_{eff}$ up to third order in the expansion of the angles $\Delta\theta$ and $\Delta\phi_1$. This is enough to give us the $\Gamma_{cusp}^\b$ up to fourth order in the angles expansion.
In Section 4 we calculate the cusp anomalous dimension in perturbation theory and up to two-loops. We argue that, up to this order, the Wilson loop expectation value in the deformed theory does not depend on $\beta$. Consequently, the cusp anomalous dimension is identical to that of $\cN=4$ sYM and its expression obtains a similar form to that of equations \eq{cuspstrong0} to \eq{BPS-cond0}. Its precise form is given in equations \eqref{cuspweak} to \eq{weakL}.\\
Combining these two results and taking into account the similarities of the $\beta$-deformed theory to the undeformed one, we conjecture that the cusp anomalous dimension of the deformed theory can be written for any value of the coupling $\lambda$ in terms of the undeformed one as
\begin{eqnarray}\label{cuspgeneral1}
\Gamma_{cusp}^{\beta}(\varphi,\Delta\theta,\Delta\phi_1,\lambda,\beta)= \Gamma_{cusp}^{N=4}(\varphi,\Delta\theta_{eff},\lambda),\,\,\,\,\forall ~\lambda~.
\end{eqnarray}
If true then it follows that there exists a BPS-like condition of the form
\begin{eqnarray}\label{BPS-cond1}
\varphi=\Delta\theta_{eff}(\Delta\theta,\Delta\phi_1,\lambda,\beta).
\end{eqnarray}
It would be nice to understand \eqref{cuspgeneral1} and \eqref{BPS-cond1} from a more general point of view possibly involving symmetries.
Our conclusion and further discussion is given in Section 5.
Our main text is supported by the Appendix A where we recover the $1/L$ conformal behavior of the quark/anti-quark potential when the two quarks are placed at a distance $L$ at the boundary of the
$AdS_5$ space in Poincare coordinates. The proportionality constant depends on the deformation parameter and can  be expressed in terms of the same function $\Delta\theta_{eff}$ introduced above (see \eqref{qqq}).

\section{The Marginally Deformed Gauge/Gravity Duality}

The type IIB supergravity solution that is dual to the
$\beta$-deformation of $\cN=4$ super Yang Mills was found
in~\cite{Lunin:2005jy}. There are several ways to obtain the $\beta$ deformed background from the original $AdS_5\times S^5$. One way is to reduce the ten dimensional theory to eight dimensions on a torus. The eight dimensional supergravity then is invariant under an $SL\prt{2,R}$ symmetry acting on the K\"{a}hler modulus $\t$ which has as a real part the $B$-field and as an imaginary the volume of the two torus. Acting with the $SL\prt{2,R}$ on the background with a deformation parameter $\gh$, the Lunin-Maldacena backgrounds can be generated.

The most convenient way, however, to obtain this deformed background is to perform a TsT deformation on the original metric. This consists of a T-duality along a $U\prt{1}$ direction, then a shift on an another $U\prt{1}$ direction because of which the deformation parameter enters and finally a T-duality along the initial  $U\prt{1}$ symmetry. Notice that the corresponding beta deformation can be performed in all dualities that
possess a $U\prt{1}\times U\prt{1}$ global symmetry, as for example in \cite{Giataganas:2010mj}.

The deformed supergravity solution \cite{Lunin:2005jy} contains the metric on $AdS_5\times {\tilde S}^5$ where ${\tilde S}^5$
is a $\beta$-deformed five-sphere. It also involves the
dilaton-axion field $\tau$ as well as the RR and NS-NS form fields.
In the string frame it reads
\bea \label{LM}
    ds^2 &=& R^2 \left[ds^2_{AdS_5}+  \sum_i \left(d \mu_i^2+ G
            \mu_i^2 d\phi_i^2\right)
        + \hat{\gamma}^2 G \mu_1^2 \mu_2^2 \mu_3^2 \big(
                \sum_i d\phi_i \big)^2
        \right]\,, \label{LM-metric}\\
    B &=& R^2 \gh \ G\ (\mu_1^2 \mu_2^2 d\phi_1 \wedge d\phi_2
        + \mu_2^2 \mu_3^2 d\phi_2 \wedge d\phi_3 + \mu_3^2
                \mu_1^2 d\phi_3 \wedge d\phi_1)\,,\\\nn
    C_2 &=& - 4 R^2 \gh\ \omega_1 \wedge (d\phi_1 + d\phi_2 + d\phi_3)\,,\\\nn
    C_4 &=& \omega_4 + 4 R^4 G\ \omega_1\wedge d\phi_1 \wedge
        d\phi_2 \wedge d\phi_3\,,\\
            e^{2\phi} &= g_s G\,, \quad g=4 \pi g_s
\eea
where  $R^4 = 4 \pi g_s N$ (in units where $\a'=1$),
\begin{equation}
    G^{-1} = 1 + \hat{\gamma}^2 (\mu_1^2 \mu_2^2 + \mu_2^2 \mu_3^2
        + \mu_3^2 \mu_1^2)\,.
\end{equation}
The parameter $\hat{\gamma}$ appearing in~\eqref{LM}
 is related to the deformation parameter $\beta$ of the gauge theory by:
\begin{equation}
    \hat{\gamma}= R^2\ \beta\,=\sqrt{\l}~ \beta.
\end{equation}
The definitions of $\o_1$ and $\o_4$, are not needed for our considerations and can be found in \cite{Lunin:2005jy}.
Furthermore we use
\be
\m_3=\sin \th \sin \a,\quad \m_1=\sin \th \cos\a,\quad \m_2=\cos\th~,\quad\mbox{with}\quad\sum_i^3 \m_i^2=1~.
\ee
The field theory dual of the aforementioned supergravity solution is obtained from the Lagrangian of ${\cal N} = 4$ sYM deformed by exactly
marginal operators \cite{Leigh:1995ep} which preserves the $\cN=1$ supersymmetry and has a global $U\prt{1}\times U\prt{1}$ symmetry.
The component Lagrangian of the $\beta$-deformed theory reads
\bea
\nonumber
&&{\cal{L}} = \, \Tr \Bigg( {1 \over 4}
F^{\mu \nu}F_{\mu \nu} +
(D^\mu \bar \Phi^i ) (D_\mu \Phi_i  )
- {g^2\over 2} [\Phi_i,\Phi_j]_{*}[\bar \Phi^i,\bar \Phi^j]_{*}
+{g^2\over 4}[\Phi_i,\bar \Phi^i][\Phi_j,\bar \Phi^j] \\
&&+ \l_{A} \s^{\mu} D_{\mu} \bar\l^A
- i g ([\l_4,\l_i]\bar \Phi^i+[\bar\l^4,\bar \l^i]\Phi_i)
+{i g\over 2}(\epsilon^{ijk}[\l_i,\l_j]_{*}\Phi_k
+\epsilon_{ijk}[\bar\l^i,\bar\l^j]_{*}\bar \Phi^k)\Bigg)~.
\nonumber \\
\label{Ldef}
\eea
Here $\Phi_i,\,\,\,i=1,2,3$ are the three complex scalars of the theory which are the lowest components of the corresponding superfields.
It can be thought as coming from the $\cN=4$ sYM by redefining the usual multiplication of the fields to a product which includes an exponential phase of the form $e^{n i \pi\beta}$ where the integer $n$ depends on the $U\prt{1}$ charge that has been assigned to the particular fields. More particularly we have introduced the $\beta$-deformed commutator of fields which is
\be
[ f_i,g_j]_{*} :=\, f_i * g_j - g_j * f_i = \,
e^{ i \pi \beta_{ij} }\, f_i g_j -\,
e^{ -i \pi \beta_{ij} }\, g_j f_i \ , \label{combeta}
\ee
and $\beta_{ij}$ is defined as
\be
\beta_{ij}=-\beta_{ji} \, , \quad
\beta_{12}=\, -\beta_{13}=\, \beta_{23} :=\, \beta \ .
\label{betaijs}
\ee

\section{$\Gamma_{cusp}$ of the $\beta$-deformed Theory at Strong Coupling}

The Wilson line which has a cusp with angle $\varphi$ on the plane, is mapped
to a quark-antiquark configuration in the cylinder, where the quarks represented by infinitely extended lines along the time direction, and are sitting at two points on the spatial $S^3$ forming an angle $\Delta\phi=\pi-\varphi$. When $\varphi=0$ the lines are at the antidiametric points of the sphere.
If we additionally place the pair of the two distinct points of the sphere, at an angle $\theta$, the configuration becomes
supersymmetric in the undeformed $\cN=4$ sYM, when the following BPS condition is satisfied, $\th^2=\varphi^2$. Our intension is to find $\Gamma_{cusp}$, as well as to find out if any BPS-like condition exists and
how this condition is  modified in the $\beta$-deformed
gauge/gravity duality. We should stress that it is not a-priori clear that such a condition should exist since the notion of 1/4 BPS observables in
${\cal N}=1$ theories is not defined.
We find indications that the relation for the $\Gamma_{cusp}$ is of the form \eqref{cuspgeneral}.

In what follows we obtain the strongly coupled BPS-like condition \eq{BPS-cond}. We show that this condition
is equivalent to the undeformed $AdS\times S$ condition, by making a definition of an effective angle of the sphere. 
To determine $\G_{cusp}$ we 
parameterize the deformed 2-sphere by the angles $\theta$ and $\phi_1$. The latter describes a $U\prt{1}$ global symmetry of the sphere which has  the following metric
\be
ds^2= R^2\prt{d\theta^2+G \sin^2\theta d\phi_1^2}~.
\ee
We take both the space-time and the world-sheet metric to be Euclidean.
Thus, the total metric of the background is
\begin{eqnarray}
ds^2= R^2(\cosh^2{\rho}\,dt^2+ d\rho^2+ \sinh^2{\rho}\,d\phi^2+ d\theta^2+ G\, sin^2{\theta} \,d\phi_1^2)~.
\end{eqnarray}
The choice of this subsector is consistent and can be verified that the equations of motion for the additional coordinates are satisfied.
The string configuration we choose is
\bea
&&t= \t, \quad \rho=\rho(\s), \quad \phi=\phi(\s)\\
&&\th=\th(\s),\quad \phi_1=\phi_1(\s)~,
\eea
for which the $G$ factor takes the form
\begin{eqnarray}\label{grs3}
G=\frac{1}{1+\gh^2 sin^2{\theta}cos^2{\theta}}~.
\end{eqnarray}
The boundary conditions for the string world-sheet are
\bea
&&\rho(\s_0)=\rho(\s_1)=\infty~,\qquad\quad~ ~\phi(\s_0)=\phi_{0,AdS}~,~ \phi(\s_1)=\phi_{1,AdS}~,\\&&
\theta(\s_0)=\theta_0,~ \theta(\s_1)=\theta_1~,\qquad
\phi_1(\s_0)=\hat{\phi_{0}},~ \phi_1(\s_1)=\hat{\phi_{1}}~,
\eea
where $\D\theta:=\theta_1-\theta_0$, $\D\phi:=\phi_{1,AdS}-\phi_{0,AdS}$, $\D\phi_1:=\hat{\phi_1}-\hat{\phi_0}$.
Using the Polyakov action
\be
S_E=\frac{\sqrt{\l}}{4 \pi }\int d\s d\t\sqrt{-\gamma} \gamma^{ab} \partial_a X^\m\partial_b X^\n g_{\m\n}~,
\ee
we find the final form of the Virasoro constraints and the equations of motion needed to be solved
\begin{eqnarray}\label{eom11}
&&\rho'^2=\cosh^2{\rho}-\frac{c^2}{\sinh^2{\rho}}-c_2~,\\
\label{eom22}
&&\sinh^2{\rho} \phi'=c~,\\\label{eom33}
&&\theta'^2=c_2-\frac{c_1^2}{4 G sin^2{\theta}}~,\\\label{eom44}
&&\phi_1'=\frac{c_1}{2 G sin^2{\theta}}~,
\end{eqnarray}
where $c,~c_1$ and $c_2$ are integration constants. The equation \eq{eom33} for $\theta$ is reduced to first order differential equation by combining the initial second order differential equation for $\theta$ obtained from the variation of the Polyakov action, with equation \eq{eom44} and forming a total derivative. The method to obtain the equation of $\theta$ is used and presented with more details in the Appendix A. The Virasoro constraint \eq{eom11} has taken its final form after the equations of motion have been used. Differentiating it with respect to $\s$ gives an equation identical to equation of motion for $\rho$.

Making the redefinition $y=\cosh\rho$ and integrating the equations \eq{eom11}, \eq{eom22}, corresponding to the equations of coordinates of the $AdS$ space we get
\bea\label{eom11b}
&& \Delta \sigma=\frac{1}{y_2}\Big( K(\frac{y_1^2}{y_2^2})-\frac{1}{\sqrt{\frac{y_1^2}{y_2^2}}}K(\frac{y_2^2}{y_1^2}) -F(-\arcsin{\frac{y_2}{y_1}}\Big|\frac{y_1^2}{y_2^2})\Big)~,\\
\nn
&&\Delta\phi = \frac{c}{y_2}\left(\frac{-i}{y_1^2-1}(K(1-\frac{y_1^2}{y_2^2})-y_1^2 \Pi(1-y_1^2\big|1-\frac{y_1^2}{y_2^2}))-\Pi(y_1^2;i \sinh^{-1}{\frac{y_2}{\sqrt{-y_1^2}}}\Big|\frac{y_1^2}{y_2^2})\right)~,\\\label{eom22b}
\eea
where $\Delta\sigma=\sigma_1-\sigma_0$, $y_2$ is the turning point, and $y_1$ is another constant given by \eq{yy2a}. The functions appear in the above expressions are the elliptic integrals of first kind $K\prt{m}$, $F\prt{n|m}$, the elliptic integral of the third kind $\P\prt{n|m}$  and the incomplete elliptic integral $\P\prt{n;k|m}${}\footnote{The definitions and conventions of the elliptic functions are the same with Mathematica.}.
Working with the integrations of the equations of \eq{eom33}, \eq{eom44} we achieve to bring the result in a relatively compact form,
by identifying the quantities which  appear repeatedly and redefining them as below
\bea\label{eom33b}
&&\Delta \sigma=-\frac{|\cos\th| \sqrt{2 \gh^2 \cos^2\th-a_1} \sqrt{\frac{2-8 c_3}{a_1 \cos^2\th}+1}~ F\left(\sin ^{-1}\left(\sqrt{\frac{8 c_3-2}{\cos^2\th a_1}}\right)|\frac{a_1}{a_2}\right)}{\gh \sqrt{\frac{4 c_3-1}{a_1}} \sqrt{\frac{-\gh^2 \cos^2\th \sin^2\th+4 \sin^2\th c_3-1}{c_3}}}\Big|_{\th_0}^{\th_0+\D\th}~,
\eea
\bea\nn
&&\D \phi_1=|\cos^3\th| \Bigg[-\frac{4 \sqrt{\frac{4 c_3-1}{a_1}} \prt{-\gh^2 \cos^2\th \sin^2\th+4 \sin^2\th c_3-1}}{\cos^4\th}-\frac{1}{2\gh^2|\cos\th|}\sqrt{-4 \gh^2+\frac{2 a_1}{\cos^2\th}}\\\nn
&&\sqrt{-2 \gh^2+\frac{a_2}{\cos^2\th}}\Bigg(-(a_2+2) F\prt{i \sinh ^{-1}\prt{\sqrt{\frac{2-8 c_3}{a_1\cos^2\th}}}\Big|\frac{a_1}{a_2}}+\\\nn
&&
y_2 E\prt{i \sinh^{-1}\prt{\sqrt{\frac{2-8 c_3}{a_1\cos^2\th}}}\Big|\frac{a_1}{a_2}}+2 \Pi \prt{\frac{a_1}{8 c_3-2};i \sinh ^{-1}\prt{\sqrt{\frac{2-8 c_3}{a_1\cos^2\th}}}\Big|\frac{a_1}{a_2}}
\Bigg)\Bigg]\\\label{eom44b}
&&\prt{4 \sqrt{\frac{4 c_3-1}{a_1}} \sqrt{-\gh^2 \cos^2\th \sin^2\th+4 \sin^2\th c_3-1}}^{-1}\Big|_{\th_0}^{\th_0+\D\th}~.
\eea
The function $E\prt{n|m}$ is the elliptic integral of second kind, and the parameters $a_1$ and $a_2$ depend on the deformation $\gh$ explicitly in the following way
\bea\label{yy2a}
&&y_1^2=\frac{1}{2}\Big(1+c_2-\sqrt{(1-c_2)^2+4 c^2} \Big)~,\qquad y_2^2=\frac{1}{2}\Big(1+c_2+\sqrt{(1-c_2)^2+4 c^2} \Big)~,\\\nn
&&a_1 = \gh^2 + 4 c_3+ \sqrt{\gh^4 + \gh^2 (4 - 8 c_3) + 16 c_3^2}~,~
a_2 = \gh^2 + 4 c_3 - \sqrt{\gh^4 + \gh^2 (4 - 8 c_3) + 16 c_3^2}~,
\eea
with $c_3=c_2/c_1^2$.

An important comment is in order. An effective angle $\Delta\theta_{eff}$ of the following form
\begin{eqnarray}\label{thetaeffective}
\Delta\theta_{eff}=\int_{\theta_0}^{\theta_0+\Delta\theta} \, d\theta \, \frac{1}{\sqrt{1- \frac{A(\theta)}{c_3}}}~,\quad\mbox{where}\quad A(\theta)=\frac{1}{4 G \sin^2\theta}~,
\end{eqnarray}
may be defined so that the integrated equation \eqref{eom33} becomes
\be\label{oopp1}
\Delta\s=\frac{\Delta\theta_{eff}}{\sqrt{c_2}}~.
\ee
This implies that the system of equations \eqref{eom11b}, \eq{eom22b} and \eq{oopp1} becomes the same as the corresponding system of equations in the undeformed theory ${\cal N}=4$ sYM but with the redefined angle $\Delta\theta_{eff}$ in the place of the angle $\Delta\theta_{S^5}$, where in the undeformed theory $\Delta\theta_{S^5}$ is the angle of the cusp in the internal space. Calculating the expression for the quark anti-quark potential in the deformed theory we see that it depends only on $y_1$ and the turning point of the string $y_2$ and not on $c_1$ and $\gh$ explicitly. It reads
\begin{eqnarray}\label{energyqq1}
E_{Q\bar{Q}}=\frac{\sqrt{\l}}{\pi}\prt{\int_{y_2}^\infty dy\frac{y^2}{\sqrt{(y^2-y_1^2)(y^2-y_2^2)}}-\int_{1}^\infty dy \frac{y}{\sqrt{y^2-1}}}~,
\end{eqnarray}
where $y=\cosh\rho$.
Notice that the constants $y_{1,2}$ include the $c_2$ dependence, where the $\gh$ dependence is hidden.

Therefore by taking into account the results of the undeformed $\cN=4$ sYM \cite{Correa:2012at,Drukker:2011za}, we conclude that the cusp anomalous dimension or equivalently the quark/anti-quark potential in the deformed theory will be of the following form
\begin{eqnarray}\label{cuspstrong}
 \Gamma_{cusp}^{\beta}(\varphi,\Delta\theta,\Delta\phi_1,\lambda,\beta)= \Gamma_{cusp}^{N=4}(\varphi,\Delta\theta_{eff},\lambda)+{\cal O}(\lambda^0)~.
\end{eqnarray}
Notice that the dependence of cusp anomalous dimension on the deformation parameter $\gh$ enters through the effective angle $\Delta\theta_{eff}$, so the results depend on $\gh$.
Furthermore, \eqref{cuspstrong} implies that,
at strong coupling at least, there exists a BPS-like condition defined by $\varphi=\Delta\theta_{eff}(\Delta\theta,\Delta\phi_1,\lambda,\beta)$.
When this condition is satisfied the cusp anomalous dimension becomes zero. Note that this condition depends explicitly on the deformation parameter and the coupling constant. This is not peculiar since there are other cases where a BPS condition depends on the coupling constant. The best known case is the dispersion relation of the giant magnon. In the section \ref{section:weak}, we will see that a similar expression to  \eq{cuspstrong} holds also at weak coupling up to two-loops. 
Therefore it is natural to conjecture that the cusp anomalous dimension in the $\beta$-deformed theory is given by
\begin{eqnarray}\label{cuspgeneral}
\Gamma_{cusp}^{\beta}(\varphi,\Delta\theta,\Delta\phi_1,\lambda,\beta)= \Gamma_{cusp}^{N=4}(\varphi,\Delta\theta_{eff},\lambda),\,\,\,\,\forall ~\lambda~
\end{eqnarray}
and depends on $\beta$. As a result, there exists a non-trivial coupling dependent function
$\Delta\theta_{eff}(\Delta\theta,\Delta\phi_1,\l,\beta)$ which
defines a BPS-like condition in the $\beta$-deformed theory
\begin{eqnarray}\label{BPS-cond}
\varphi=\Delta\theta_{eff}(\Delta\theta,\Delta\phi_1,\lambda,\beta).
\end{eqnarray}
This conjecture is further supported by the following facts: a) the field content of the $\cN=4$ sYM and the $\beta$ deformed theory is identical, b) the expectation values of several observables, e.g. the $1/4$ BPS-like Wilson loops, remain undeformed,
plus c) that both theories are exactly conformal. We expect that the condition \eqref{BPS-cond} holds for any value of the coupling.
The expression for $\Delta\theta_{eff}$ is determined in the next subsection at the strong coupling regime, and  in Section \ref{section:weak} at weak coupling regime.

In practice one can determine the function $\Delta\theta_{eff}$ by inverting the equation \eqref{eom44b}, to find $c_3$ in terms of the angles $\theta_0$, $\Delta\theta$, $\Delta\phi_1$. Then the equation \eqref{thetaeffective} should be integrated and the expression for $c_3$ should be substituted to the results to obtain $\Delta\theta_{eff}$. In the next subsection we work along this way and obtain $\Delta\theta_{eff}$ up to third-order in the angles expansion.

It is interesting to notice that the string solutions we have found here corresponding to Wilson loops in the $\beta$-deformed $\cN=4$ sYM,
depend on the parameter $\gh$ and reduce to the undeformed  solutions smoothly as $\gh\rightarrow 0$. This characteristic is nice and is different to  non-supersymmetric string solutions in other beta deformed dualities as the Sasaki-Einstein ones \cite{Giataganas:2010mj,Giataganas:2011zza}, which may not reduce to acceptable undeformed solutions \cite{Giataganas:2009an,Giataganas:2009dr}.
We close this section by noticing, that the equation \eq{cuspstrong} seems to be  quite generic for spaces of the form $AdS_5\times M$. When the solutions to the equations of motion exist, in order for the Virasoro constraint to be satisfied they need to combine in such a way, to give a version of \eq{cuspstrong}. Furthermore, we should mention that
the non-trivial content of the $\G_{cusp}$ expression \eqref{cuspgeneral} is that given the function $\D \theta_{eff}$ all mixed terms involving both the $AdS_5$ and $S^5$ angles like $\phi^2 \Delta\theta^2$ are uniquely determined.

\subsection{$\Delta\theta_{eff}$ up to Third-order in the Angle Expansion}\label{section:f1eq}
In what follows, we solve equation \eq{eom44b} for $c_3$ up to third order in the small angle expansion. This is enough to determine the generalized cusp anomalous dimension in the deformed theory up to fourth order in the angle expansion.
To this end, we expand the integrand of
\begin{eqnarray}\label{intphi1}
\Delta\phi_1=\int_{\theta_0}^{\theta_0+\Delta\theta} \, d\theta \, 2 A(\theta)\frac{1}{\sqrt{c_3- A(\theta)}}\,\,\,\, ~,
\end{eqnarray}
up to first order in $\theta-\theta_0$ and integrate the resulting expression to get
\begin{eqnarray}\nn
\Delta\phi_1=&&\frac{2A(\theta_0) \Delta\theta}{\sqrt{c_3- A(\theta_0)}}+ \Delta\theta^2 A'\prt{\th_0} \Big(
\frac{1}{\sqrt{c_3- A(\theta_0)}}+\frac{A(\theta_0)}{2(c_3- A(\theta_0))^{3/2}} \Big)
\\&& +\frac{\partial^2}{\partial \th_0^2}\prt{\frac{2A(\theta_0) }{\sqrt{c_3- A(\theta_0)}}}\frac{\Delta\theta^3}{6} +{\mathcal O}(\Delta\theta^4)~,\label{zeq}
\end{eqnarray}
where $A'\prt{\th_0}:=\partial A(\theta_0)/\partial\theta_0$. The above equation is solved analytically for $c_3$ in Appendix \ref{app:c3}. The next step is to expand the equation \eq{thetaeffective} for $\D\th_{eff}$ to the third order and get
\begin{eqnarray}\label{theff}
\Delta\theta_{eff}^{str}=\frac{\Delta\theta}{\sqrt{1- \frac{A(\theta_0)}{c_3}}}+\frac{\D\theta^2}{4(1- \frac{A(\theta_0)}{c_3})^{3/2}}\frac{1}{c_3}\frac{\partial A(\theta_0)}{\partial \theta_0}+ \frac{\D\theta^3}{6}\frac{\partial^2}{\partial \theta_0^2 }\frac{1}{\sqrt{1- \frac{A(\theta_0)}{c_3}}} + {\mathcal O}(\Delta\theta^4)~.
\end{eqnarray}
Plugging in \eqref{theff} the value of $c_3$ obtained in Appendix B we obtain the solution up to third order
\bea\label{theffinal}\nn
\Delta\theta_{eff}^{str}&=&\sqrt{\Delta\theta^2+G_0 \sin^2\theta_0 \Delta\phi_1^2}+\frac{\Delta\th^2}{4}
 \frac{G_0\prt{1+\prt{2 G_0-1}\cos 2\th_0} \sec\th_0}{ \sqrt{\omega\prt{G_0+\omega\csc^2\th_0}}}\\&&+ \Delta \theta^3 f_1+ {\mathcal O}(\Delta\theta^4) ~,
\eea
where $\omega=\Delta\theta^2/\Delta\phi_1^2$, $G_0=G\prt{\th_0}$ and $f_1$ is a long expression that is presented in Appendix \ref{app:f1}. One can see that the first term is the infinitesimal distance on the deformed two sphere. By localizing the string at a fixed azimuthal angle $\D\phi_1=0$ the result becomes undeformed $\Delta \theta_{eff}=\Delta\theta$. Our result of \eq{theffinal} is adequate to give the cusp anomalous dimension up to fourth order in the angle expansion, by employing the known expression for the strongly coupled cusp anomalous dimension $\cN=4$ sYM, namely
\be
\frac{4\pi \G_{cusp}^{\b}(\varphi,\Delta\theta,\Delta\phi_1,\lambda,\beta)}{\sqrt{\l}} =\frac{\Delta\theta_{eff}^2-\phi^2}{\pi}-\frac{1}{8 \pi^3}\prt{\Delta\theta_{eff}^2-\phi^2}\prt{\Delta\theta_{eff}^2-5\phi^2}+{\mathcal O}(\Delta\theta^6) ~.
\ee

\section{$\Gamma_{cusp}$ of the $\beta$-deformed Theory at Weak Coupling} \label{section:weak}
\begin{figure*}[!ht]
\centerline{\includegraphics[width=0.9\textwidth]{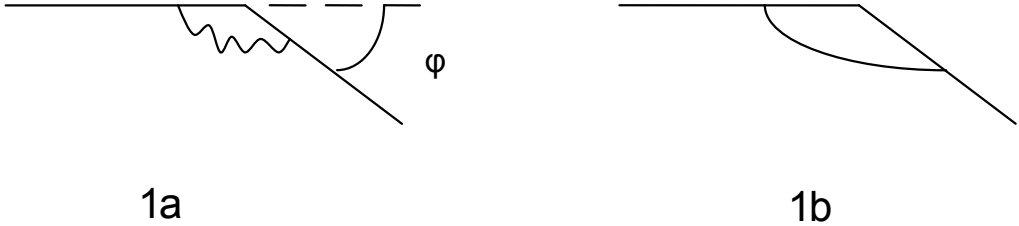}}
\caption{  1-loop Feynman diagrams (1a-1b) contributing to the cusp anomalous dimension. Wiggly lines denote gluon propagators while
solid lines denote the scalar propagators. No confusion should be caused by the fact that we use the same kind of line (solid) to depict both the scalar propagator and  the Wilson line.}\label{fig:fig1}
\end{figure*}
\begin{figure*}[!ht]
\centerline{\includegraphics[width=.9\textwidth]{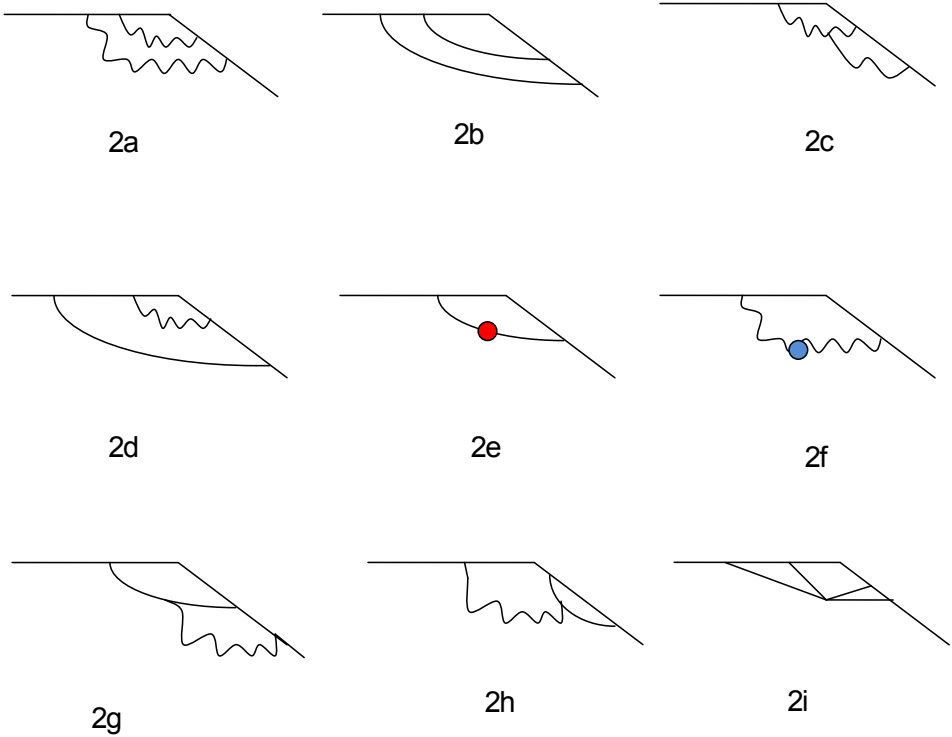}}
\caption{\label{StringDiagrms} 2-loop Feynman diagrams (2a-2h) that contribute to the cusp anomalous dimension. Wiggly lines denote gluon propagators while solid lines denote the scalar propagators. The red blob denotes the 1-loop correction to the scalar propagator while the blue one
the 1-loop correction to the gluon propagator.
All 2-loop diagrams do not depend on $\beta$.
The last diagram (2i) which introduces $\beta$-dependence contributes at 3-loops. There are other similar diagrams to the above, for example with two gluon propagators emitted from the horizontal line of the Wilson loop as in 2c, that contribute to the cusp anomalous dimension but do not introduce $\beta$ dependence. }\label{fig:fig2}
\end{figure*}
In this section we use the following definition the Wilson loop operator in the $\beta$-deformed ${\cal N}=1$ sYM theory
\begin{eqnarray}\label{Wilsonloop}
&&W[C]=\frac{1}{N}\tr P \exp \prt{i g \oint (A_{\mu} \dot{x^{\mu}}\prt{\t}-i \dot{y}^i\prt{\t} \phi_i)d\t},\\\nn
&&\quad\mbox{with}\quad \dot{y}_i=n^i |\dot{x}|,\quad\sum_{i=1}^6 n_i^2=1~.
\end{eqnarray}
Our choice can be justified from several points of view. It has been found that the $1/4$ BPS-like Wilson loop in the $\beta$ deformed theory in the strong coupling limit has undeformed expectation value \cite{Chu:2007pb}. This has also been proved in the case of higher representation Wilson loops in \cite{Imeroni:2006rb}. Therefore, it is natural to think that the strong coupling results of the expectation values indicate an undeformed Wilson loop operator. A further indication to this direction is that the UV cancelation divergence condition $\dot{x}^2=\dot{y}^2$ derived rigorously by the Legendre transformation, has been proved to remain undeformed \cite{Chu:2007pb,Chu:2008xg}. In the weak coupling regime, it has been shown that such a Wilson loop operator has a finite value at least up to order $\prt{g^2 N}^2$.

The contour of integration $C$ of the \eq{Wilsonloop} is depicted in Figure 1. It consists of two straight lines forming a cusp of angle $\varphi$ in Euclidean space-time. Furthermore, there is a angle in the internal space. One of the semi-infinite lines is associated with the point A: $\vec{n}_1=(\sin\theta_0 \cos\hat{\phi}_0,\sin\theta_0 \sin\hat{\phi_0},\cos\theta_0)$ of the deformed  two-sphere $\tilde {S}^2\subset \tilde {S}^5$ while the other semi-infinite line is associated with the point B: $\vec{n}_2=\prt{\sin(\theta_0+\Delta\theta) \cos(\hat{\phi_0}+\Delta\phi_1),\sin(\theta_0+\Delta\theta) \sin(\hat{\phi_0}+\Delta\phi_1),\cos(\theta_0+\Delta\theta)}$, where $\Delta\theta$ and $\Delta\phi_1$ are the difference of the angles defining the point A and the point B in the internal space.

We are now in position to evaluate the one and two-loop contribution to the expectation value of \eqref{Wilsonloop}. The corresponding diagrams
 are depicted in Figures 1 and 2, respectively. By inspecting the Lagrangian density \eqref{Ldef} one can easily convince himself that all one and two-loop diagrams are independent of the deformation parameter $\beta$. The reason is that any $\beta$-dependence can only be introduced through the 4-scalar interaction vertex (see Figure \ref{fig:fig2}). More particularly, the 1-loop self-energy of the gluon and scalar propagators are independent of $\beta$ \cite{Chu:2007pb}. This claim can also be verified as follows. The scalar 1-loop self-energy (Figure 2e) receives contribution from two diagrams. One where the scalar field emits a gluon and then reabsorbs it. This diagram has no $\beta$ dependence as can be easily seen from the second term of \eqref{Ldef}. A second contribution to the  scalar self-energy is coming from the diagram where the scalar splits in two fermions through a Yukawa interaction. The fermions propagate and annihilate through a second Yukawa interaction to give the scalar. The relevant terms in the Lagrangian \eqref{Ldef} are the four last terms. This diagram either does not depend on $\beta$, if one of the intermediate fermions is $\lambda_4$, or it involves two Yukawa vertices each of which
has a  $\beta$ dependence that is the inverse of the other. This is so because one Yukawa vertex involves $\lambda_i,\, \, \lambda_j$ while the second  involves ${\bar \lambda}_i,\, \, {\bar \lambda}_j$
and  $\lambda$ and  ${\bar \lambda}$ have opposite $U(1)$ charges. As a result, the scalar 1-loop self-energy does not depend on  $\beta$.

Similar arguments can be used to show that the gluon 1-loop self-energy (Figure 2f) is $\beta$ independent. In this case there are four diagrams contributing. One where the intermediate state is a pair of two fermions, one where the intermediate state is a pair of two gluons,  one where the intermediate state is a pair of two scalars and one where the intermediate state is a pair of  ghost fields.  None of  these diagrams  depends on the deformation parameter  $\beta$.
We conclude that up to 2-loops the expectation value of the Wilson loop is identical to that of the parent theory ${\cal N}=4$ sYM. However, we expect that from 3-loops and beyond the result will depend explicitly on the deformation parameter.

To summarize, the Wilson loop expectation value of Figure 1 and as a result the expression of the cusp anomalous dimension, will be the same in the marginally deformed and undeformed theories up to the two-loop order. In particular, this means that we write
 \begin{eqnarray}\label{cuspweak}
 \Gamma_{cusp}^{\beta}(\varphi,\Delta\theta, \Delta\phi_1,\lambda,\beta) =\Gamma_{cusp}^{N=4}(\varphi, \Delta\theta_{eff},\lambda)+{\cal O}(\lambda^3)~,
 \end{eqnarray}
 with the BPS-like condition in the deformed theory being
 \be \varphi=\Delta\theta_{eff}+{\cal O}(\lambda^2)~,
 \ee
 where the angle $\Delta\theta_{eff}$ between the points A and B on the sphere is given in terms of $\Delta\theta$ and $\Delta\phi_1$ by $\cos\Delta\theta_{eff}=\vec{n}_1 \cdot \vec{n}_2$.
 Thus we have determined the weak coupling expression for $\Delta\theta_{eff}$ to be {}\footnote{Let us mention that the weak coupling expansion of
 $\Delta\theta_{eff}$ is of the form $\Delta\theta_{eff}(\Delta\theta,\Delta\phi_1,\theta_0,\lambda,\beta)=
 \Delta\theta_{eff}^{(0)}(\Delta\theta,\Delta\phi_1,\theta_0,\beta)$ $ +g^2\Delta\theta_{eff}^{(1)}(\Delta\theta,\Delta\phi_1,\theta_0,\beta)$ $
 +g^4\Delta\theta_{eff}^{(2)}(\Delta\theta,\Delta\phi_1,\theta_0,\beta)$.
 The fact that the 2-loop cusp dimension is the same as in ${\cal N}=4$ sYM up to 2-loops implies that the 1-loop expansion coefficient
 $\Delta\theta_{eff}^{(1)}$ is zero.}
\be\label{weakL}
\Delta\theta_{eff}(\Delta\theta,\Delta\phi_1,\theta_0,\lambda,\beta) =arc\cos(\vec{n}_1\cdot \vec{n}_2)+{\cal O}(\l^2)~.
\ee

\section{Discussion}

In this paper we have calculated the cusp anomalous dimension of the Wilson loop in $\beta$ deformed gauge/gravity duality. An interesting feature of this theory is that several BPS-like quantities, have an undeformed expectation value in the strong coupling limit, for example as\cite{Chu:2007pb,Imeroni:2006rb}. This fact enables us to make a conjecture for the from of the cusp anomalous dimensions for every $\lambda$.

In the gravity side we achieve to partially map the system of equations to the undeformed system by redefining an angle $\Delta\theta_{eff}$ which depends on the t'Hooft constant, and the angles $\theta$ and $\phi_1$ of the internal space. This mapping is enough to relate the  cusp anomalous dimension in Lunin-Maldacena background with the one in the undeformed $AdS$ space as
\begin{eqnarray}\label{cuspgeneralg}
\Gamma_{cusp}^{\beta}(\varphi,\Delta\theta,\Delta\phi_1,\lambda,\beta)= \Gamma_{cusp}^{N=4}(\varphi,\Delta\theta_{eff},\lambda),\,\,\,\,\lambda>>1~,
\end{eqnarray}
for all $\beta$, where the $\D\theta_{eff}\prt{\beta,\l,\D\th,\D\phi_1}$ has been found. It also implies that at strong coupling we obtain a BPS-like condition in the deformed theory being $\varphi=\Delta\theta_{eff}\prt{\beta,\l}+{\cal O}(\lambda^0)$, where the angle $\Delta\theta_{eff}$ is given by \eq{theffinal}
\footnote {Additionally, it seems the form of $\Gamma_{cusp}$ in the strong coupling regime is similar for backgrounds of the generic form $AdS\times M$. Nevertheless, for each case the equations of motion in the internal space $M$ need to be solved explicitly.} .

In the field theory side, by choosing appropriately the Wilson loop operator, we show that up to two loops there is no $\beta$ dependence, although for higher loops it might appear. Consequently, one can write the expression \eq{cuspweak} for the cusp anomalous dimension of the deformed theory. This expression is similar to \eqref{cuspgeneralg}, where now $\Delta\theta_{eff}$ is given by \eqref{weakL}.

Therefore, taking into account the properties of the $\beta$ deformed theory, it is natural to conjecture that the cusp anomalous dimension of the deformed theory takes a form similar to \eq{cuspgeneralg}, i.e. see the equations \eq{cuspgeneral} and \eq{BPS-cond}, for any value of the coupling constant $\lambda$.   The function $\Delta\theta_{eff}$ which enters in the definition of the BPS-like condition $\varphi=\Delta\theta_{eff}\prt{\beta,\l}$ is determined in the strong and weak coupling limits in sections 3 and 4.  It would be interesting to check our conjecture through a derivation of the TBA equations for $\Gamma_{cusp}$ along the lines of \cite{Correa:2012hh,Drukker:2012de}.

Notice that by localizing the string along the $U(1)$ direction, i.e. $\Delta \phi_1=0$ in the deformed sphere, the solution becomes identical to the undeformed one, that is $\Delta\theta_{eff}=\Delta\theta_{S^5}$ . It is also interesting that our results extend in a straightforward way to the multi-$\beta$ deformed theory, realized by deforming the 5-sphere with three TsT deformations and resulting to a theory with no supersymmetry \cite{Frolov:2005dj}. This implies that the role of supersymmetry in the cusp anomalous dimensions may not be the leading one.

Moreover, since in the real $\beta$ deformed theory the $AdS_5$ and $\tilde{S}^5$ part are separated in the metric, the dimensionless bremsstrahlung function $B\prt{\lambda,N}$ defined by
\be
\Delta E=2\pi B\prt{\lambda,N}\int dt\dot{v}^2
\ee
remains undeformed in the strongly coupled $\beta$ deformed theory. Therefore the energy loss of the quark moving in the conformal real $\beta$ deformed theory is not affected by the deformation. This is in agreement with the observation of \cite{Correa:2012at} that the value of $B\prt{\lambda,N}$ is expected to be the same for any conformal field theory.
\newline~\\
\textbf{Acknowledgments} \newline
This work of G. G. is partly supported by the General
Secretariat for Research and Technology of Greece and from the European Regional Development Fund (NSRF 2007-13 ACTION, $KPH\Pi I \Sigma$).
The research of D. G. is implemented under the "ARISTEIA" action of the "operational programme education and lifelong learning"
 and is co-funded by the European Social Fund (ESF) and National Resources.  Part of this work was done while D.G. was supported by a SPARC and earlier by a Claude Leon postdoctoral fellowship.

\startappendix
\section{Analysis in Poincare Coordinates and $1/L$ Behavior}
In this section we make evident the conformal $1/L$ behavior of the Wilson loop corresponding to the energy of the Q\={Q} in the $\beta$ deformed duality with string extended along the deform sphere. We work in the Poincare coordinates, so the $AdS$ part becomes\footnote{From now on we set $R=1$ which is equivalent to absorb it in the constants and also set $\e=1$.}
\be
ds^2=u^2(dx_0^2+d\vec{x}^2)+\frac{ du^2}{u^2}+d\tilde{\Omega}_5~,
\ee
where $\tilde{\Omega}_5$ is the metric of the deformed sphere \eq{LM}.
The string configuration reads
\bea
&&x_0= \t, \quad x_1=x_1(\s)\\
&&\a=\a(\s),\quad \th=\th(\s),\quad \phi_1=\phi_1(\s),\quad \phi_2=\phi_2(\s)~.
\eea
We use the Polyakov action and we see that, it is consistent to set
\be
\phi_2=constant~.
\ee
The equation of motion for $\a$ gives
\be
\partial_\a G_{\phi_1\phi_1}=0\Rightarrow\a=0,~~\mbox{meaning}~~\m_3=0~.
\ee
The equation for $x_1$, $u$ and the Virasoro constrain are
\bea\label{eomx1}
&&x_1'=\frac{c }{2 u^2}~,\\
\label{eomu0}
&&u(1+x_1'^2)+\frac{ u'^2}{u^3}-\frac{1}{u^2}u''=0~,\\
\label{vic2}
&&u^2(-1+x_1'{}^2)+ 1(\th'^2 +G \sin^2\th \phi_1'{}^2)+\frac{1}{u^2}u'^2=0~.
\eea
The equation for $\theta$ gives a second order differential equation
while the equation for $\phi$ can be straightforward integrated as
\bea\label{eomth0}
&&2\th''=\partial_\th(G \sin^2{\th})\phi_1'{}^2~,\\\label{eomphi}
&&\phi_1'=\frac{c_1}{2  G \sin^2\th}~.
\eea
Combining the above equations we get a total derivative in the equation of motion for
$\theta$
\be\label{eomth3}
\th'^2=c_2-\frac{c_1^2}{4 G \sin^2\th }~.
\ee
where $c_2$ is a constant comes from the integration of $\th'$. The equation can brought to the integral form
\be\label{sth}
\Delta\s=\int_{\th_0}^{\th_1}\frac{d\theta}{\sqrt{c_2-\frac{c_1^2}{4 G\sin^2\th}}}~.
\ee
Similarly the equation \eq{eomphi} gives
\begin{eqnarray}\label{intphi}
\Delta\phi_1=\int_{\theta_0}^{\theta_0+\Delta\theta} \, d\theta \, 2 A(\theta)\frac{1}{\sqrt{\frac{c_2}{c_1^2}- A(\theta)}},\,\,\,\,\, A(\theta)=\frac{1}{4G\sin^2\theta} ~,
\end{eqnarray}
Substituting the equations of motion to \eq{eomx1}, \eq{eomphi}, \eq{eomth3} to Virasoro constrain \eq{vic2} becomes a
differential equation for $u$
\be\label{uprime}
u'^2= u^4-c_2 u^2-\frac{c^2}{4}
\ee
which gives
\be\label{su}
\Delta\s= 2\int_{0}^{u_0}\frac{du}{\sqrt{u^4-c_2 u^2-\frac{c^2}{4}}}~.
\ee
Using \eq{uprime} the equation for $x_1$ \eq{eomx1} can be brought to the integrable form
\be\label{xu}
L=2\int_{0}^{u_0}\frac{  c}{ 2 u^2\sqrt{\e^2 u^4-c_2 u^2-\frac{c^2}{4}}} du
\ee
Therefore the equations \eq{sth}, \eq{intphi}, \eq{su} and \eq{xu}, need to be integrated. The equations of motion for $\theta$ and $\phi_1$ are the same to \eq{eom33} and \eq{eom44}, we define the same angle $\Delta \theta_{eff}$ \eq{thetaeffective}, and thinking along the same lines of the main text arguments we conclude that the energy of Q\={Q} is similar to the undeformed case, with $\D\theta_{eff}$ instead of $\D\theta$ of the $\cN=4$ sYM derived in  \cite{Maldacena:1998im}
\bea
&&\frac{L}{2 } =\frac{1}{u_0 } \sqrt{1-c_4^2 } I_1(c_4) ~,\\\label{lc4}
&&\quad\frac{ \Delta \theta_{eff}}{2} = c_4 I_2(c_4) ~,
\eea
where
\bea
&&I_1(c_4) = \frac{1}{(1 - c_4^2) \sqrt{ 2 - c_4^2 } }\left[
(2 -c_4^2 )E\left(\frac{ \pi}{ 2 }, \sqrt{\frac{1-c_4^2}{ 2 - c_4^2 }}\right) -
F\left( \frac{\pi}{2},  \sqrt{\frac{1-c_4^2}{ 2 - c_4^2 }}\right) \right] ~,\\
&&I_2(c_4) = \frac{ 1}{ \sqrt{2 - c_4^2 } }
F\left( \frac{ \pi}{ 2}, \sqrt{\frac{1-c_4^2} {2 - c_4^2 }}\right) ~.
\eea
The $u_0$ is the turning point of $u$ specified by setting $u'$ to zero from equation \eq{uprime}, and $c_4$ is an integration constant. Equation
\eq{intphi} contains the deformation parameter $\gh$ and when integrated is given from  equation \eq{eom44b}. Therefore the energy is given by
\be\label{qqq}
E_{QQ}=\frac{2\sqrt{2 \lambda }}{\pi L}\prt{1-c_4^2}^{3/2}I_1^2(c_4)
\ee
where $c_4$ depends on the deformation parameter through \eq{lc4}. Other special string solutions corresponding to the energy of the dipole of quarks were found in \cite{Hernandez:2005xd,Hernandez:2005zx}, where also the results depend on the deformation $\gh$ parameter.

\section{Solution for $c_3$ of the equation \eq{zeq}}\label{app:c3}
In this section we present the solution of the equation \eq{zeq} for $c_3$. Defining $y:=c_3- A(\theta_0)$ and taking the square of
\eqref{zeq} we obtain
\begin{eqnarray}\label{yeq}\nn
\Delta\phi_1^2 ~y^5=&&4 A^2(\theta_0) y^4 \Delta\theta^2 +4 A(\theta_0)y^3 A'\prt{\th_0}\Big( y+\frac{1}{2}A(\theta_0)\Big)\Delta\theta^3 \\\nn
&&+\Bigg[A'\prt{\th_0}^2
y^2\prt{ y+\frac{1}{2}A(\theta_0)}^2+\frac{2A(\theta_0) y^2}{3}\Bigg( A''(\theta_0)y\prt{2 y+A(\theta_0)}
\\&& +A'\prt{\th_0}^2\prt{2 y  +\frac{3}{2}A(\theta_0)} \Bigg)\Bigg]\Delta\theta^4~.
\end{eqnarray}
One can solve \eqref{yeq} perturbatively. The zeroth order solution is obtained by ignoring in \eqref{yeq} all terms of order $\Delta\theta^3$ and higher and reads
\begin{eqnarray}\label{yzero}
y^{(0)}=\frac{4 A^2(\theta_0)\Delta\theta^2 }{\Delta\phi_1^2}.
\end{eqnarray}
Then substituting
\begin{eqnarray}\label{yansatz}
y=y^{(0)}+y^{(1)}, {\rm where}\,\,\,y^{(0)}\sim \Delta\theta^0,\,\,y^{(1)}\sim \Delta\theta^1
\end{eqnarray}
in \eqref{yeq} we obtain a quadratic equation in $y^{(1)}$ of the standard form $a\,(y^{(1)})^2+b \,y^{(1)}+c=0 $ which can be easily solved to give
\begin{eqnarray}\label{y1sol}
y^{(1)}=\frac{1}{4a} \prt{\frac{b_1^2-4 a c}{ b_0}- \prt{\frac{b_1-2 a c_1}{b_0}}^2},
\end{eqnarray}
where
\begin{eqnarray}\label{yansatz1}
&&a=-4 (y^{(0)})^3,\,\,\,\,b_0=-(y^{(0)})^4,\,\,\,\,b_1= \,\frac{A'\prt{\th_0}}{A(\theta_0)} (y^{(0)})^3
\prt{4y^{(0)}+\frac{3}{2} A(\theta_0) }\Delta\theta\,\,\,\,\,\nonumber \\ \nn
&&c_1= \frac{A'\prt{\th_0}}{A(\theta_0)}(y^{(0)})^4 \prt{y_0+\frac{A(\theta_0)}{2}} \Delta\theta ,~~~ c_2= \frac{(y^{(0)})^3}{4 A(\theta_0)^2}\Bigg(
A'\prt{\th_0}^2\Big(y^{(0)}+\frac{A(\theta_0)}{2} \Big)^2\\&& ~\vspace{5cm}~+\frac{2}{3} A(\theta_0)\prt{A''\prt{\th_0}y_0\prt{2 y_0+A(\theta_0)}+A'(\theta_0)^2\prt{2y_0+\frac{3A(\theta_0)}{2}}}\Bigg) \Delta\theta^2,\,\,\,\,\nonumber \\&&{\rm with}\,\,\,\,
b:=b_0+b_1~,\,\,\,\,c:=c_1+c_2~.
\end{eqnarray}
Thus $c_3=y_0+y_1+A(\theta_0)$ has been specified.
Here we should mention that when substituting \eqref{yansatz} in \eqref{yeq} we have kept all
terms up to fourth order in the small quantities $\Delta\theta$ and $\Delta\phi_1$ which is
enough to determine the expression for $y$ up to quadratic order in the small angles $\Delta\theta$ and $\Delta\phi_1$.
In particular, this implies that one can safely ignore the term proportional to $ \Delta\phi_1^2 (y^{(1)})^3$ since this is of fifth order.
In this way the cubic in $y$ equation \eqref{yeq} becomes quadratic in $y^{(1)}$.
A last comment is in order. Should we wish to keep terms of higher order in the angle
expansion we would need to keep more terms in the expansion
of \eqref{intphi1} and the subsequent analysis.

\section{Third order mathematical expression in \eq{theffinal}} \label{app:f1}
In this section we present the analytic form of the long expression $f_1$ appearing in the equation \eq{theffinal}. We have found
\bea\nn
&&f_1:=G_0 \Bigg[\sin ^5\th_0 \bigg(\big((G_0+\o \csc ^2\th_0) \big(\big(48 \o^5 \csc ^8\th_0 \sec ^2\th_0 ((2 G_0-1) \cos 2 \th_0+1)^2
\\&&\nn(-G_0 \cos 2 \th_0+G_0+4 \o)^2\big)(G_0+\o \csc ^2\th_0)^{-1}-\big(\sin ^{14}\th_0 (16 G_0^8 \o^4 \csc ^{16}\th_0 \\\nn&&\big(1024 G_0^3 \o \cot ^2 2 \th_0+\frac{1}{4} (G_0-1) \cos 2 \th_0 \csc ^4\th_0 (87 G_0^2 \cos 4 \th_0+261 G_0^2
\\\nn&&-4 G_0 (87 G_0+256 \o) \cos 2 \th_0+1024 G_0 \o+1920 \o^2)+\frac{3}{8} (G_0-1)^2 \cos ^2 2 \th_0 \csc ^4\th_0
\\\nn&&\sec ^2\th_0(-116 G_0^2 \cos 2 \th_0+29 G_0^2 \cos 4 \th_0+87 G_0^2+640 \o^2)-2048 G_0^2 \o \cot ^2 2 \th_0
\\\nn&&+87 G_0^2 \cos ^2 \th_0
+288 G_0 \o \cot ^2\th_0+1024 G_0 \o \cot ^2 2 \th_0+16 G_0 \o \csc ^2 \th_0 +32 \o^2 \csc ^4 \th_0
\\\nn&&+304 \o^2 \cot ^2\th_0 \csc ^2\th_0\big)
-3 \sec ^2\th_0 ((2 G_0-1) \cos 2 \th_0+1)^2 \big(-96 G_0^9-256 G_0^8 \o \csc ^2\th_0
\\\nn&&+G_0 \o^4 \csc ^{16}\th_0+2 \o^5 \csc ^{18}\th_0\big)^2)\big)G_0^{-8}\big)\big)\o^{-4}+36 G_0 \csc ^6\th_0 \sec ^2\th_0 ((2 G_0-1) \cos 2 \th_0+1)^2
\\\nn&&(-G_0 cos 2 \th_0+G_0+4 \o)^2\bigg)-24 \csc \th_0 \sec ^2\th_0 ((2 G_0-1) \cos 2 \th_0+1)^2 (-G_0 \cos 2 \th_0
\\\nn&&+G_0+4 \o) (-3 G_0 \cos 2 \th_0+3 G_0+4 \o) (G_0+\o \csc ^2\th_0)+64 \sin \th_0 (G_0+\o \csc ^2\th_0)^2
\\\nn&& \big(3 G_0 \sin ^22 \th_0+12 (G_0-1)^2 G_0 \cos ^2 2 \th_0 \tan ^2 \th_0+24 (G_0-1) G_0 \sin ^2 \th_0 \cos 2 \th_0
\\&&+8 \o \cos ^2\th_0
+4 \o\big)\Bigg]\big(1536 \o^{5/2} (G_0+\o \csc ^2 \th_0 )^{3/2}\big)^{-1}
\eea

\bibliographystyle{nb}
\bibliography{botany}

\begin{thebibliography}{10}
\ifx\href\asklfhas\newcommand{\href}[2]{#2}\fi
\ifx\arxivref\asklfhas\newcommand{\arxivref}[2]{\href{http://arxiv.org/abs/#1}%
{#2}}\fi
\ifx\doiref\asklfhas\newcommand{\doiref}[2]{\href{http://dx.doi.org/#1}{#2}}\fi
\raggedright
\small
\parskip 0pt

\bibitem{Maldacena:1998re}
J.~M.~Maldacena,
\textit{``The large N limit of superconformal field theories and
  supergravity''},
\textsf{Adv.~Theor.~Math.~Phys.~2,~231~(1998)},
\texttt{\arxivref{hep-th/9711200}{hep-th/9711200}}.

\bibitem{Witten:1998qj}
E.~Witten,
\textit{``Anti-de Sitter space and holography''},
\textsf{Adv.~Theor.~Math.~Phys.~2,~253~(1998)},
\texttt{\arxivref{hep-th/9802150}{hep-th/9802150}}.

\bibitem{Beisert:2010jr}
N.~Beisert, C.~Ahn, L.~F.~Alday, Z.~Bajnok, J.~M.~Drummond et~al.,
\textit{``{Review of AdS/CFT Integrability: An Overview}''},
\textsf{\doiref{10.1007/s11005-011-0529-2}{Lett.Math.Phys.~99,~3~(2012)}},
\texttt{\arxivref{1012.3982}{arxiv:1012.3982}}.

\bibitem{Koch:2011hb}
R.~d.~M.~Koch, M.~Dessein, D.~Giataganas and C.~Mathwin,
\textit{``{Giant Graviton Oscillators}''},
\textsf{\doiref{10.1007/JHEP10(2011)009}{JHEP~1110,~009~(2011)}},
\texttt{\arxivref{1108.2761}{arxiv:1108.2761}}.

\bibitem{Carlson:2011hy}
W.~Carlson, R.~d.~M.~Koch and H.~Lin,
\textit{``{Nonplanar Integrability}''},
\textsf{\doiref{10.1007/JHEP03(2011)105}{JHEP~1103,~105~(2011)}},
\texttt{\arxivref{1101.5404}{arxiv:1101.5404}}.

\bibitem{Escobedo:2010xs}
J.~Escobedo, N.~Gromov, A.~Sever and P.~Vieira,
\textit{``{Tailoring Three-Point Functions and Integrability}''},
\texttt{\arxivref{1012.2475}{arxiv:1012.2475}}.

\bibitem{Escobedo:2011xw}
J.~Escobedo, N.~Gromov, A.~Sever and P.~Vieira,
\textit{``{Tailoring Three-Point Functions and Integrability II. Weak/strong
  coupling match}''},
\texttt{\arxivref{1104.5501}{arxiv:1104.5501}}.

\bibitem{Gromov:2012uv}
N.~Gromov and P.~Vieira,
\textit{``{Tailoring Three-Point Functions and Integrability IV.
  Theta-morphism}''},
\texttt{\arxivref{1205.5288}{arxiv:1205.5288}}.

\bibitem{Georgiou:2012zj}
G.~Georgiou, V.~Gili, A.~Grossardt and J.~Plefka,
\textit{``{Three-point functions in planar N=4 super Yang-Mills Theory for
  scalar operators up to length five at the one-loop order}''},
\textsf{\doiref{10.1007/JHEP04(2012)038}{JHEP~1204,~038~(2012)}},
\texttt{\arxivref{1201.0992}{arxiv:1201.0992}}.

\bibitem{Georgiou:2011qk}
G.~Georgiou,
\textit{``{SL(2) sector: weak/strong coupling agreement of three-point
  correlators}''},
\textsf{\doiref{10.1007/JHEP09(2011)132}{JHEP~1109,~132~(2011)}},
\texttt{\arxivref{1107.1850}{arxiv:1107.1850}}.

\bibitem{Georgiou:2010an}
G.~Georgiou,
\textit{``{Two and three-point correlators of operators dual to folded string
  solutions at strong coupling}''},
\textsf{\doiref{10.1007/JHEP02(2011)046}{JHEP~1102,~046~(2011)}},
\texttt{\arxivref{1011.5181}{arxiv:1011.5181}}.

\bibitem{Georgiou:2008vk}
G.~Georgiou, V.~L.~Gili and R.~Russo,
\textit{``{Operator Mixing and the AdS/CFT correspondence}''},
\textsf{\doiref{10.1088/1126-6708/2009/01/082}{JHEP~0901,~082~(2009)}},
\texttt{\arxivref{0810.0499}{arxiv:0810.0499}}.

\bibitem{Leigh:1995ep}
R.~G.~Leigh and M.~J.~Strassler,
\textit{``Exactly marginal operators and duality in four-dimensional
  {$\mathcal{N}=\mathord{}$1} supersymmetric gauge theory''},
\textsf{\doiref{10.1016/0550-3213(95)00261-P}{Nucl.~Phys.~B447,~95~(1995)}},
\texttt{\arxivref{hep-th/9503121}{hep-th/9503121}}.

\bibitem{Lunin:2005jy}
O.~Lunin and J.~Maldacena,
\textit{``Deforming field theories with U(1)$\mathord{}\times\mathord{}$U(1)
  global symmetry and their gravity duals''},
\textsf{\doiref{10.1088/1126-6708/2005/05/033}{JHEP~0505,~033~(2005)}},
\texttt{\arxivref{hep-th/0502086}{hep-th/0502086}}.

\bibitem{Frolov:2005dj}
S.~Frolov,
\textit{``{Lax pair for strings in Lunin-Maldacena background}''},
\textsf{\doiref{10.1088/1126-6708/2005/05/069}{JHEP~0505,~069~(2005)}},
\texttt{\arxivref{hep-th/0503201}{hep-th/0503201}}.

\bibitem{Freedman:2005cg}
D.~Z.~Freedman and U.~Gursoy,
\textit{``{Comments on the beta-deformed N=4 SYM theory}''},
\textsf{\doiref{10.1088/1126-6708/2005/11/042}{JHEP~0511,~042~(2005)}},
\texttt{\arxivref{hep-th/0506128}{hep-th/0506128}}.

\bibitem{Mauri:2005pa}
A.~Mauri, S.~Penati, A.~Santambrogio and D.~Zanon,
\textit{``{Exact results in planar N=1 superconformal Yang-Mills theory}''},
\textsf{\doiref{10.1088/1126-6708/2005/11/024}{JHEP~0511,~024~(2005)}},
\texttt{\arxivref{hep-th/0507282}{hep-th/0507282}}.

\bibitem{Mauri:2006uw}
A.~Mauri, S.~Penati, M.~Pirrone, A.~Santambrogio and D.~Zanon,
\textit{``{On the perturbative chiral ring for marginally deformed N=4 SYM
  theories}''},
\textsf{\doiref{10.1088/1126-6708/2006/08/072}{JHEP~0608,~072~(2006)}},
\texttt{\arxivref{hep-th/0605145}{hep-th/0605145}}.

\bibitem{Fiamberti:2008sm}
F.~Fiamberti, A.~Santambrogio, C.~Sieg and D.~Zanon,
\textit{``{Finite-size effects in the superconformal beta-deformed N=4 SYM}''},
\textsf{\doiref{10.1088/1126-6708/2008/08/057}{JHEP~0808,~057~(2008)}},
\texttt{\arxivref{0806.2103}{arxiv:0806.2103}}.

\bibitem{Khoze:2005nd}
V.~V.~Khoze,
\textit{``{Amplitudes in the beta-deformed conformal Yang-Mills}''},
\textsf{\doiref{10.1088/1126-6708/2006/02/040}{JHEP~0602,~040~(2006)}},
\texttt{\arxivref{hep-th/0512194}{hep-th/0512194}}.

\bibitem{Chu:2006ae}
C.-S.~Chu, G.~Georgiou and V.~V.~Khoze,
\textit{``{Magnons, classical strings and beta-deformations}''},
\textsf{\doiref{10.1088/1126-6708/2006/11/093}{JHEP~0611,~093~(2006)}},
\texttt{\arxivref{hep-th/0606220}{hep-th/0606220}}.

\bibitem{Bykov:2008bj}
D.~V.~Bykov and S.~Frolov,
\textit{``{Giant magnons in TsT-transformed AdS(5) x S**5}''},
\textsf{\doiref{10.1088/1126-6708/2008/07/071}{JHEP~0807,~071~(2008)}},
\texttt{\arxivref{0805.1070}{arxiv:0805.1070}}.

\bibitem{Georgiou:2006np}
G.~Georgiou and V.~V.~Khoze,
\textit{``{Instanton calculations in the beta-deformed AdS/CFT
  correspondence}''},
\textsf{\doiref{10.1088/1126-6708/2006/04/049}{JHEP~0604,~049~(2006)}},
\texttt{\arxivref{hep-th/0602141}{hep-th/0602141}}.

\bibitem{Durnford:2006nb}
C.~Durnford, G.~Georgiou and V.~V.~Khoze,
\textit{``{Instanton test of non-supersymmetric deformations of the AdS(5) x
  S**5}''},
\textsf{\doiref{10.1088/1126-6708/2006/09/005}{JHEP~0609,~005~(2006)}},
\texttt{\arxivref{hep-th/0606111}{hep-th/0606111}}.

\bibitem{Chu:2007pb}
C.-S.~Chu and D.~Giataganas,
\textit{``{Near BPS Wilson Loop in beta-deformed Theories}''},
\textsf{\doiref{10.1088/1126-6708/2007/10/108}{JHEP~0710,~108~(2007)}},
\texttt{\arxivref{0708.0797}{arxiv:0708.0797}}.

\bibitem{Imeroni:2006rb}
E.~Imeroni and A.~Naqvi,
\textit{``{Giants and loops in beta-deformed theories}''},
\textsf{\doiref{10.1088/1126-6708/2007/03/034}{JHEP~0703,~034~(2007)}},
\texttt{\arxivref{hep-th/0612032}{hep-th/0612032}}.

\bibitem{Berenstein:2004ys}
D.~Berenstein and S.~A.~Cherkis,
\textit{``{Deformations of N=4 SYM and integrable spin chain models}''},
\textsf{\doiref{10.1016/j.nuclphysb.2004.09.005}{Nucl.Phys.~B702,~49~(2004)}},
\texttt{\arxivref{hep-th/0405215}{hep-th/0405215}}.

\bibitem{Beisert:2005if}
N.~Beisert and R.~Roiban,
\textit{``{Beauty and the twist: The Bethe ansatz for twisted N=4 SYM}''},
\textsf{\doiref{10.1088/1126-6708/2005/08/039}{JHEP~0508,~039~(2005)}},
\texttt{\arxivref{hep-th/0505187}{hep-th/0505187}}.

\bibitem{Frolov:2005ty}
S.~Frolov, R.~Roiban and A.~A.~Tseytlin,
\textit{``{Gauge-string duality for superconformal deformations of N=4 super
  Yang-Mills theory}''},
\textsf{\doiref{10.1088/1126-6708/2005/07/045}{JHEP~0507,~045~(2005)}},
\texttt{\arxivref{hep-th/0503192}{hep-th/0503192}}.

\bibitem{Gromov:2010dy}
N.~Gromov and F.~Levkovich-Maslyuk,
\textit{``{Y-system and $\beta$-deformed N=4 Super-Yang-Mills}''},
\textsf{\doiref{10.1088/1751-8113/44/1/015402}{J.Phys.~A44,~015402~(2011)}},
\texttt{\arxivref{1006.5438}{arxiv:1006.5438}}.

\bibitem{Arutyunov:2010gu}
G.~Arutyunov, M.~de~Leeuw and S.~J.~van~Tongeren,
\textit{``{Twisting the Mirror TBA}''},
\textsf{\doiref{10.1007/JHEP02(2011)025}{JHEP~1102,~025~(2011)}},
\texttt{\arxivref{1009.4118}{arxiv:1009.4118}}.

\bibitem{deLeeuw:2010ed}
M.~de~Leeuw and T.~Lukowski,
\textit{``{Twist operators in N=4 beta-deformed theory}''},
\textsf{\doiref{10.1007/JHEP04(2011)084}{JHEP~1104,~084~(2011)}},
\texttt{\arxivref{1012.3725}{arxiv:1012.3725}}.

\bibitem{Zoubos:2010kh}
K.~Zoubos,
\textit{``{Review of AdS/CFT Integrability, Chapter IV.2: Deformations,
  Orbifolds and Open Boundaries}''},
\textsf{\doiref{10.1007/s11005-011-0515-8}{Lett.Math.Phys.~99,~375~(2012)}},
\texttt{\arxivref{1012.3998}{arxiv:1012.3998}}.

\bibitem{Ahn:2010ws}
C.~Ahn, Z.~Bajnok, D.~Bombardelli and R.~I.~Nepomechie,
\textit{``{Twisted Bethe equations from a twisted S-matrix}''},
\textsf{\doiref{10.1007/JHEP02(2011)027}{JHEP~1102,~027~(2011)}},
\texttt{\arxivref{1010.3229}{arxiv:1010.3229}}.

\bibitem{Ahn:2012hs}
C.~Ahn, M.~Kim and B.-H.~Lee,
\textit{``{Worldsheet S-matrix of beta-deformed SYM}''},
\textsf{\doiref{10.1016/j.physletb.2013.01.047}{Phys.Lett.~B719,~458~(2013)}},
\texttt{\arxivref{1211.4506}{arxiv:1211.4506}}.

\bibitem{Maldacena:1998im}
J.~M.~Maldacena,
\textit{``{Wilson loops in large N field theories}''},
\textsf{\doiref{10.1103/PhysRevLett.80.4859}{Phys.Rev.Lett.~80,~4859~(1998)}},
\texttt{\arxivref{hep-th/9803002}{hep-th/9803002}}.

\bibitem{Zarembo:2002an}
K.~Zarembo,
\textit{``{Supersymmetric Wilson loops}''},
\textsf{\doiref{10.1016/S0550-3213(02)00693-4}{Nucl.Phys.~B643,~157~(2002)}},
\texttt{\arxivref{hep-th/0205160}{hep-th/0205160}}.

\bibitem{Drukker:1999zq}
N.~Drukker, D.~J.~Gross and H.~Ooguri,
\textit{``{Wilson loops and minimal surfaces}''},
\textsf{\doiref{10.1103/PhysRevD.60.125006}{Phys.~Rev.~D60,~125006~(1999)}},
\texttt{\arxivref{hep-th/9904191}{hep-th/9904191}}.

\bibitem{Correa:2012at}
D.~Correa, J.~Henn, J.~Maldacena and A.~Sever,
\textit{``{An exact formula for the radiation of a moving quark in N=4 super
  Yang Mills}''},
\textsf{\doiref{10.1007/JHEP06(2012)048}{JHEP~1206,~048~(2012)}},
\texttt{\arxivref{1202.4455}{arxiv:1202.4455}}.

\bibitem{Correa:2012hh}
D.~Correa, J.~Maldacena and A.~Sever,
\textit{``{The quark anti-quark potential and the cusp anomalous dimension from
  a TBA equation}''},
\textsf{\doiref{10.1007/JHEP08(2012)134}{JHEP~1208,~134~(2012)}},
\texttt{\arxivref{1203.1913}{arxiv:1203.1913}}.

\bibitem{Drukker:2012de}
N.~Drukker,
\textit{``{Integrable Wilson loops}''},
\texttt{\arxivref{1203.1617}{arxiv:1203.1617}}.

\bibitem{Makeenko:2006ds}
Y.~Makeenko, P.~Olesen and G.~W.~Semenoff,
\textit{``{Cusped SYM Wilson loop at two loops and beyond}''},
\textsf{\doiref{10.1016/j.nuclphysb.2006.05.002}{Nucl.Phys.~B748,~170~(2006)}},
\texttt{\arxivref{hep-th/0602100}{hep-th/0602100}}.

\bibitem{Correa:2012nk}
D.~Correa, J.~Henn, J.~Maldacena and A.~Sever,
\textit{``{The cusp anomalous dimension at three loops and beyond}''},
\textsf{\doiref{10.1007/JHEP05(2012)098}{JHEP~1205,~098~(2012)}},
\texttt{\arxivref{1203.1019}{arxiv:1203.1019}}.

\bibitem{Drukker:2011za}
N.~Drukker and V.~Forini,
\textit{``{Generalized quark-antiquark potential at weak and strong
  coupling}''},
\textsf{\doiref{10.1007/JHEP06(2011)131}{JHEP~1106,~131~(2011)}},
\texttt{\arxivref{1105.5144}{arxiv:1105.5144}}.

\bibitem{Forini:2010ek}
V.~Forini,
\textit{``{Quark-antiquark potential in AdS at one loop}''},
\textsf{\doiref{10.1007/JHEP11(2010)079}{JHEP~1011,~079~(2010)}},
\texttt{\arxivref{1009.3939}{arxiv:1009.3939}}.

\bibitem{Giataganas:2010mj}
D.~Giataganas,
\textit{``{Semiclassical strings in marginally deformed toric AdS/CFT}''},
\textsf{\doiref{10.1007/JHEP12(2011)051}{JHEP~1112,~051~(2011)}},
\texttt{\arxivref{1010.1502}{arxiv:1010.1502}}.

\bibitem{Giataganas:2011zza}
D.~Giataganas,
\textit{``{String solutions in Sasaki-Einstein manifolds and their marginally
  deformed versions}''},
\textsf{\doiref{10.1016/j.nuclphysbps.2011.04.163}{Nucl.Phys.Proc.Suppl.~216,~%
227~(2011)}}.

\bibitem{Giataganas:2009an}
D.~Giataganas,
\textit{``{On the non-BPS string solutions in Sasaki-Einstein gauge / gravity
  duality}''},
\textsf{\doiref{10.1007/JHEP06(2010)016}{JHEP~1006,~016~(2010)}},
\texttt{\arxivref{0912.3624}{arxiv:0912.3624}}.

\bibitem{Giataganas:2009dr}
D.~Giataganas,
\textit{``{Semi-classical Strings in Sasaki-Einstein Manifolds}''},
\textsf{\doiref{10.1088/1126-6708/2009/10/087}{JHEP~0910,~087~(2009)}},
\texttt{\arxivref{0904.3125}{arxiv:0904.3125}}.

\bibitem{Chu:2008xg}
C.-S.~Chu and D.~Giataganas,
\textit{``{UV-divergences of Wilson Loops for Gauge/Gravity Duality}''},
\textsf{\doiref{10.1088/1126-6708/2008/12/103}{JHEP~0812,~103~(2008)}},
\texttt{\arxivref{0810.5729}{arxiv:0810.5729}}.

\bibitem{Hernandez:2005xd}
R.~Hernandez, K.~Sfetsos and D.~Zoakos,
\textit{``{Gravity duals for the Coulomb branch of marginally deformed N=4
  Yang-Mills}''},
\textsf{\doiref{10.1088/1126-6708/2006/03/069}{JHEP~0603,~069~(2006)}},
\texttt{\arxivref{hep-th/0510132}{hep-th/0510132}}.

\bibitem{Hernandez:2005zx}
R.~Hernandez, K.~Sfetsos and D.~Zoakos,
\textit{``{On supersymmetry and other properties of a class of marginally
  deformed backgrounds}''},
\textsf{\doiref{10.1002/prop.200510294}{Fortsch.Phys.~54,~407~(2006)}},
\texttt{\arxivref{hep-th/0512158}{hep-th/0512158}}.

\end{thebibliography}

\end{document}